  \documentclass[structabstract]{aa}  
\usepackage{graphicx}
\usepackage{txfonts}
\usepackage{natbib}
\usepackage{longtable}

\bibpunct[]{(}{)}{;}{a}{}{,}
\begin{document}
   \title{Laboratory spectroscopy of 1,2-propanediol at millimeter and submillimeter 
          wavelengths\thanks{Full Table~4 and Tables~5$-$6 as well as full text file 
          {\tt aG'g.txt} and text files {\tt gG'a.txt} and {\tt g'G'g.txt} are only 
          available in electronic form at the CDS via anonymous ftp to 
          cdsarc.u-strasbg.fr (130.79.128.5) or via http://cdsweb.u-strasbg.fr/cgi-bin/qcat?J/A+A/}}

   \author{J.-B. Bossa\inst{1}   
           \and
           M.~H. Ordu\inst{2}
           \and
           H.~S.~P. M{\"u}ller\inst{2}
           \and
           F. Lewen\inst{2}
           \and
           S. Schlemmer\inst{2}}

   \institute{Sackler Laboratory for Astrophysics, Leiden Observatory, Leiden University, 
              P.O. Box 9513, 2300 RA Leiden, The Netherlands. 
    \and I.~Physikalisches Institut, Universit{\"a}t zu K{\"o}ln,
              Z{\"u}lpicher Str. 77, 50937 K{\"o}ln, Germany\\
              \email{hspm@ph1.uni-koeln.de}
         }

   \date{Received 1 June 2014 / Accepted 14 August 2014}

  \abstract
{Ethanediol is one of the largest complex organic molecules detected in space thus far. It has been found 
in different types of molecular clouds. The two propanediol isomers are the next larger diols. Hence, 
they are viable candidates to be searched for in space.}
{We wish to provide sufficiently large and accurate sets of spectroscopic parameters of 1,2-propanediol 
to facilitate searches for this molecule at millimeter and longer submillimeter wavelengths.}
{We recorded rotational spectra of 1,2-propanediol in three wide frequency windows between 38 
and 400~GHz.}
{We made extensive assignments for the three lowest energy conformers to yield spectroscopic parameters 
up to eighth order of angular momentum.}
{Our present data will be helpful for identifying 1,2-propanediol at moderate submillimeter or longer 
wavelengths with radio telescope arrays such as ALMA, NOEMA, or EVLA. In particular, its detection 
with ALMA in sources, in which ethanediol was detected, appears to be promising.}

\titlerunning{Millimeter-wave spectroscopy of 1,2-propanediol}

\maketitle
\hyphenation{iden-ti-fied}
\hyphenation{spec-tro-sco-py}
\hyphenation{For-schungs-ge-mein-schaft}
\hyphenation{ex-tra-po-lation}

%

\section{Introduction}
\label{intro}

The presence of molecular complexity in space has attracted considerable interest in the field 
of astrochemistry because of the question of the origin and evolution of life in the Universe. 
Laboratory spectroscopic data in the millimeter and submillimeter domains for the unique identification 
of large complex molecules are needed to interpret the ongoing spectral surveys of star-forming 
regions and young stellar objects.

Ethanediol, (CH$_2$OH)$_2$, also known as ethylene glycol, is one of the largest complex organic 
molecules detected in space to date, see, for example, the Cologne Database for Molecular Spectroscopy 
(CDMS) \citep{CDMS_1,CDMS_2}.\footnote{Internet address: http://www.astro.uni-koeln.de/cdms/molecules} 
Ethanediol was detected first in the massive and luminous Galactic center source Sagittarius~B2(N), 
Large Molecule Heimat (Sgr~B2(N-LMH) for short) \citep{det_eglyc_2002}. There is also strong 
evidence of ethanediol in three less-evolved molecular clouds in the Galactic center 
\citep{cold_GC_mols_2_2008}. Very recently, it was also detected in the hot corinos associated 
with the class~0 protostars NGC~1333-IRAS2A \cite{eglyc_etc_in_NGC-1333-IRAS2A_2014} and, 
tentatively, IRAS~16293-2422B \citep{glycald_ALMA_2013}. Finally, ethanediol was also found 
to be abundant in the outflows of comet Hale-Bopp \citep{eglyc_Hale-Bopp_2004}.

Even though ethanediol may exist in several distinct conformations, only two of them have 
been identified in the laboratory by using rotational spectroscopy 
\citep{aGg'-eglyc_1995,gGg'-eglyc_2001,aGg'-eglyc_2003,gGg'-eglyc_2004}, 
with the higher energy conformer estimated to be 2.5~kJ~mol$^{-1}$ ($\sim$300~K) above 
the lowest one. Therefore, astronomical detections of  ethanediol reported in the literature 
thus far only refer to the lowest energy conformer. The rotational spectra of both conformers 
display strong rotation-tunneling interaction caused by two equivalent minima on the potential 
energy surfaces. Molecular beam Fourier transform microwave spectroscopy (MB-FTMW) and 
microwave-microwave double resonance were required to achieve initial assignments 
\citep{aGg'-eglyc_1995,gGg'-eglyc_2001}, which paved the way for later experimental investigations 
at millimeter and submillimeter wavelengths \citep{aGg'-eglyc_2003,gGg'-eglyc_2004}.

Investigating propanediol constitutes the next step toward understanding the molecular complexity 
in space since it is a structural analog of ethanediol by replacing a hydrogen atom that is bound 
to a carbon atom with a methyl group. Moreover, comparing the relative abundance of ethanediol and 
its derivatives (e.g., propanediol) can provide insights into the formation routes of theses molecules 
and the chemical evolution of objects in which they are detected. However, care is needed since 
column densities of molecules tend to decrease with increasing complexity.

Two stable isomers of propanediol are known to date, 1,2-propanediol (CH$_3$CHOHCH$_2$OH) with the 
OH groups at two adjacent carbon atoms, and 1,3-propanediol (CH$_2$OHCH$_2$CH$_2$OH) with the 
OH groups at the outer carbon atoms. The conformational landscape of both isomers has been 
thoroughly investigated quite recently by using both FTMW spectroscopy and quantum chemical 
calculations \citep{1-2-PD_div_conf_FTMW_2009,1-3-PD_FTMW_2009}. Only two conformers were found 
experimentally for 1,3-propanediol, which both displayed tunneling caused by two equivalent 
minima on the potential energy surfaces \citep{1-3-PD_FTMW_2009}. Other conformers were calculated 
to be at least 600~K higher in energy, except for one that was calculated to be about 400~K higher in energy, 
but with a rather small dipole moment. The lowest energy conformer was studied before by using free-jet 
absorption microwave spectroscopy \citep{1-3-PD_rot_1995}. Very recently, Smirnov et al. (2013) 
performed extensive millimeter measurements, analyzed the rotation-tunneling interaction in the spectra 
of both conformers, and provided the spectroscopic basis to search for 1,3-propanediol in space 
by radio astronomy.

For 1,2-propanediol, seven conformers were identified using FTMW spectroscopy, three of which are 
relatively low and close in energy \citep{1-2-PD_div_conf_FTMW_2009}. Lovas et al. (2009)
also carried out quantum-chemical calculations on a variety of conformers, which included determining 
the relative energies, quartic centrifugal distortion parameters, and dipole moment components; 
experimental dipole values were determined for three conformers. The lowest energy conformer 
as well as two slightly higher energy forms had been studied to some extent before 
\citep{1-2-PD_lowest_conf_FTMW_2002,1-2-PD_rot_1981}. None of the conformers displayed rotation-tunneling 
splitting. Interestingly, rotational transitions of the lowest energy conformer were used very recently 
to demonstrate that FTMW spectroscopy can be used to determine the enantiomeric composition of a chiral 
molecule \citep{enatiomeric_composition_FTMW_2013}. Three-wave mixing was employed to create phase 
differences between the two enantiomers that were detected by FTMW.

We investigated the rotational spectra of the three lowest energy conformers of 1,2-propanediol at 
millimeter and submillimeter wavelengths to permit searching for them in space  by radio astronomical means.


\section{Experimental details}
\label{exptl}

The present experimental conditions are quite similar to those employed for the lower 
frequency measurements (up to 230~GHz) of \textit{n}-butyl cyanide \cite{n-BuCN_rot_2012}. 
Spectra were recorded in three frequency bands (38$-$70, 200$-$230, and 297$-$400~GHz). 
The fundamental frequency sources were computer-controlled sweep synthesizers that were referenced to a rubidium atomic clock. An Agilent E8257D sweeper was used as a 
direct frequency source to record lines up to 70~GHz. The RF output from a microwave 
generator (Rohde \& Schwarz SMF~100A; maximum frequency 43~GHz) was multiplied up to 
the desired frequency with cascaded multipliers from Virginia Diodes, Inc. (VDI). Factors 
of 16 and 21 were employed for the bands starting at 200 and 297~GHz. 
Both synthesizers enable quasi-continuous tuning in freely adjustable frequency increments 
of typically some kilohertz. DC-biased room-temperature Schottky diodes were used as detectors. 
The detector signal was coupled to a lock-in amplifier for phase-sensitive detection. 
Frequency modulation was used to reduce baseline effects; demodulation was carried out at 
$2f$, resulting in a line shape approximating the second derivative of a Gaussian.

The millimeter or submillimeter beam was formed by a standard gain horn antenna in combination 
with an HDPE lens for low-loss coupling of the beam to the 7~m long Pyrex absorption cell with 
100~mm inner diameter. The cell windows were made of PTFE and tilted by 10 degrees to reduce 
baseline effects even more. A double-pass absorption scheme was used, which
extended the absorption 
path length to 14~m.

The vapor pressure of 1,2-propanediol is rather low at room temperature, $\sim$19~Pa~\cite{1-2-PD_vapor_2009}. 
To achieve a sufficient and reasonably stable vapor pressure, measurements were carried out 
under slow flow conditions. In addition, the sample container, the needle valve, with which the flow 
of 1,2-propanediol was controlled at the inlet side, and the absorption cell were heated to about 
55$^{\rm o}$C (328~K). The sample pressure was between 1.0 and 1.5~Pa.

After the integration time was optimized, which depends on the desired signal-to-noise level, 
step size (20, 32, and 90~kHz at lower, medium, and higher frequencies) and 
scan widths, our spectrometer setup allowed for full-band sweeps with a typical speed of about 
1 to 4 weeks per 100~GHz because the lines were generally weak, even more so at lower frequencies.


\section{Conformational landscape and spectroscopic properties of 1,2-propanediol}
\label{conformations}

\begin{figure*}
\centering
  \includegraphics[width=18cm]{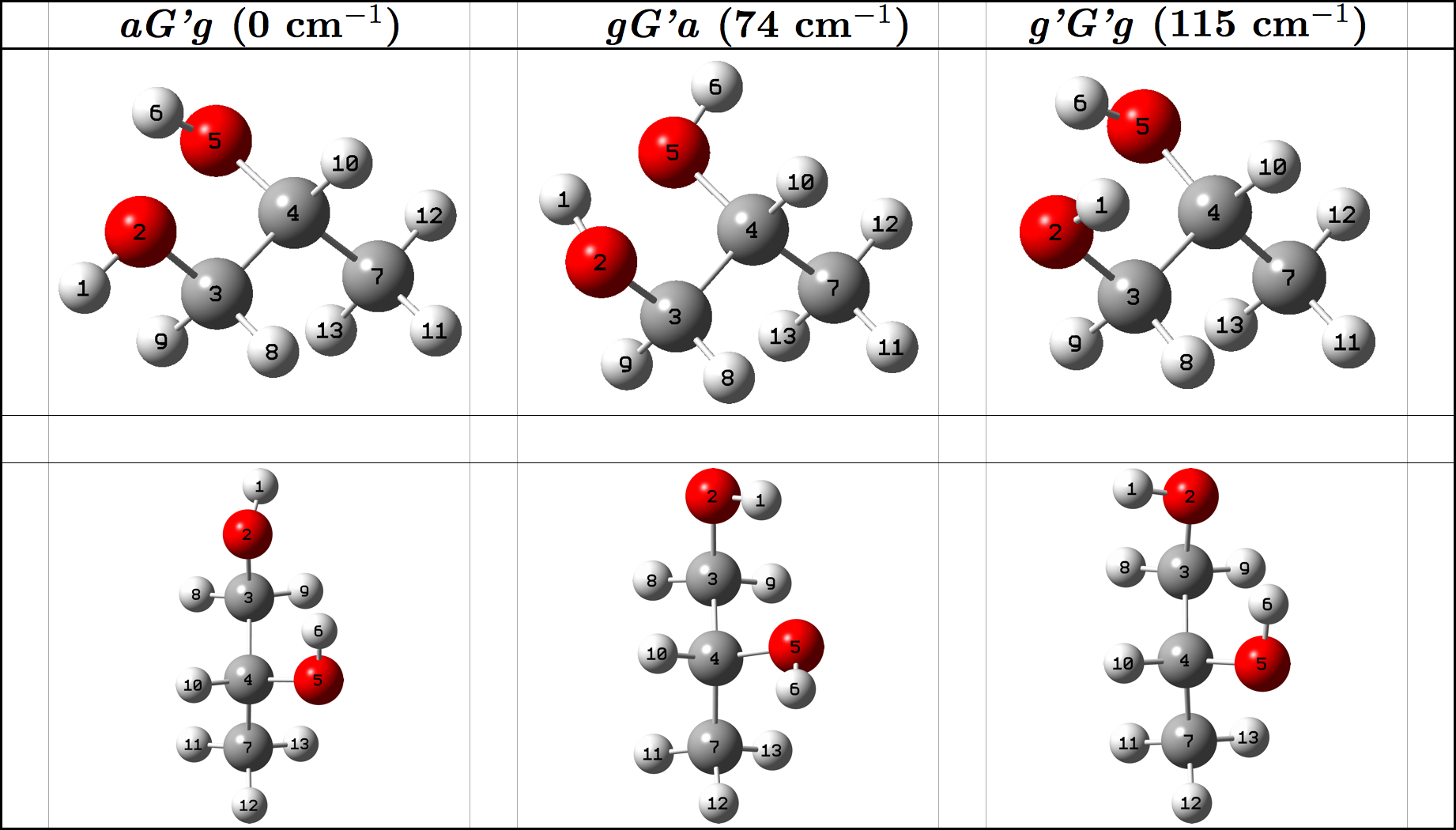} 
    \caption{Predicted structures and relative energies for the three lowest energy conformers of 
             1,2-propanediol. The C and O atoms are indicated by gray and red spheres. 
             The H atoms are indicated by smaller light-gray spheres. Atom labels and structural parameters 
             are taken from Lovas et al. (2009). Upper panel: skewed view from the side 
             of the carbon chain; lower panel: view from above the O atoms.}
\label{fig1}
\end{figure*}


The molecule 1,2-propanediol may exist in up to 27 different conformers; chirality is ignored. It is usually 
assumed that different enantiomers (conformations that behave as mirror images) have the same spectroscopic 
parameters. The assumption may be not entirely correct, but to date there is no experimental evidence for 
distinguishable spectroscopic parameters of enantiomers. Quantum chemical calculations suggest that any such 
differences would require very heavy atoms in a molecule. The conformers of 1,2-propanediol 
can be distiguished by the torsional angles of the HOCC atoms involving the outer OH group of the OCCO atoms, 
and of the CCOH atoms with the OH group at the central C atom. The OCCO torsional angle also determines 
the position of the methyl group with respect to the rest of the molecule. This angle can be about 
180$^{\rm o}$, +60$^{\rm o}$, or $-$60$^{\rm o}$, designated by $A$, $G$, and $G'$, respectively, with $A$ 
for $anti$ (or $trans$), and $G$ for $gauche$. $T$ instead of $A$ was used in Lovas et al. (2009). 
The two torsional angles involving the OH groups are designated analogously by lower case characters.

Lovas et al. (2009) used MP2 quantum chemical calculations with basis sets of triple zeta quality to 
investigate structures and energetics of the 12 1,2-propanediol conformers found to be lowest in energy in earlier 
calculations with smaller basis sets \citep{1-2-PD_ai_1989}. The first nine conformers from V{\'a}zquez et al. (1989) 
were also stable in the more recent calculations, whereas the two conformers next in energy were transformed 
into other conformers in the structure optimization process. Dipole moment components as well as quartic 
centrifugal distortion parameters were determined for the ten remaining conformers \citep{1-2-PD_div_conf_FTMW_2009}. 
Lovas et al. (2009) recorded rotational spectra of 1,2-propanediol between 6.4 and 26.0~GHz using two types of FTMW 
spectrometers and were able to assign transition frequencies to seven different conformers based on calculated 
and observed spectroscopic parameters and on the relative magnitudes of the dipole moment components. 
The conformer assignments are quite secure, in particular for the five conformers lowest in energy. 
The calculated energy ordering of these conformers appears to agree with experimental intensities 
when collisional cooling is considered, especially for the three lowest energy conformers.

The predicted structures and relative energies for the three lowest energy conformers are depicted in Fig.~\ref{fig1}.
The lowest energy conformer is $aG'g$, in agreement with an earlier FTMW study \citep{1-2-PD_lowest_conf_FTMW_2002}. 
Two other conformers, $gG'a$ and $g'Ga$, found in an even earlier millimeter wave study \citep{1-2-PD_rot_1981}, 
are the second and, probably, the fourth lowest energy conformers, calculated to be 74 and 212~cm$^{-1}$ higher. 
The third lowest conformer is $g'G'g$, 115~cm$^{-1}$ higher in energy than the lowest one.

Lovas et al. (2009) also determined dipole moment components experimentally for the $aG'g$, $gG'a,$ 
and $g'Ga$ conformers. We present in Table~\ref{tab1} the values of the two lowest energy conformers and the 
calculated components for $g'G'g$.


\begin{table}
\caption{Experimental and theoretical (in bold text) dipole moments (in Debye) for the three lowest energy 
         conformers$^a$ of 1,2-propanediol \citep{1-2-PD_div_conf_FTMW_2009}.}  
\label{tab1}      
\centering                                      
\begin{tabular}{ l l l l}          
\hline\hline                       
Conformer & $\mu_a$ & $\mu_b$ &  $\mu_c$  \\    
\hline                                   
\textit{aG'g} & 1.201 & 1.916 & 0.365 \\      
\textit{gG'a} & 2.496 & 0.309  & 0.45 \\
 \textit{\textbf{g'G'g}} & \textbf{0.41}  & \textbf{1.86}  &  \textbf{1.51}\\
\hline                                             
\end{tabular}\\[2pt]

$^a$ Conformers \textit{aG'g} and \textit{gG'a} were designated 
as \textit{tG'g} and \textit{gG't} in Lovas et al. (2009).\\

\end{table}


The three lowest energy conformers of 1,2-propanediol are fairly asymmetric top molecules with a Ray's asymmetry 
parameter $\kappa = (2B-A-C)/(A-C)$ not close to the prolate symmetric top limit of $-$1 \citep{Gordy_Cook_1984}. 
The $\kappa$ values for $aG'g$, $gG'a$, and $g'G'g$ conformers are $-$0.7062, $-$0.6900, and $-$0.7132, respectively.


\section{Results}
\label{results}

With 13 atoms, 1,2-propanediol is a comparatively heavy molecule with relatively small rotational constants, 
which, in turn, lead to a dense rotational spectrum even at low energies. The fairly large number of conformers 
as well as the larger number of usually lower lying vibrational states compared with lighter molecules increase 
the line density even more.

Lovas et al. (2009) identified more than 140 transitions between 6.4 and 26.0~GHz for 
the three lowest 1,2-propanediol conformers. Transitions obeying all three types of selection rules 
were recorded for each of the conformer. Internal rotation splitting of the outer methyl group was not 
resolved with either FTMW spectrometer. Therefore, we did not expect to resolve it either because of the 
larger line widths in our experiments.

We used Pickett's SPFIT and SPCAT programs \citep{Pickett1991} for fitting and predicting spectra of the 
1,2-propanediol conformers. Watson's $S$ reduction of the rotational Hamiltonian \citep{Watson1977} 
was used here. Predictions based on the previous works 
\citep{1-2-PD_lowest_conf_FTMW_2002,1-2-PD_div_conf_FTMW_2009} allowed us to easily assign stronger 
$R$-branch transitions with similar or slightly higher $J$ and $K_a$ quantum numbers in the 
38$-$70~GHz region. Using these transitions to improve the predictions, other assignments 
could be made until the signal-to-noise ratio was too low to determine the transition frequency 
with reasonable uncertainty. We were only able to assign transitions pertaining to the strong dipole 
moment components ($\geq$~0.5~D, see Table~\ref{tab1}; see also Table~\ref{statistics} 
for statistics on our results and final fits). 
Transitions with $J$ up to 42 and $K_a$ up to 13 were assigned in this region for the lowest energy 
$aG'g$ conformer. Similar quantum numbers were reached for the third lowest conformer $g'G'g$. 
For both conformers, $b$-type $R$- and $Q$-branch transitions were observed. In addition, 
$a$-type $R$-branch transitions were observed for the \textit{aG'g} conformer and $c$-type $R$- 
and $Q$-branch transitions for the \textit{g'G'g} conformer. Fewer assignments, all pertaining 
to $a$-type selection rules, were made for the $gG'a$ conformer because of the small $b$- and 
$c$-dipole moment components. On the other hand, we were able
to assign several weaker $Q$-branch transitions 
with $\Delta K_a = 2$ and 0 in addition to the stronger $R$-branch transitions because of the relatively 
large $a$-dipole moment component.


\begin{table}
\begin{center}
\caption{Total number of transitions and other statistical information of our
         1,2-propanediol data sets.}
\label{statistics}
\renewcommand{\arraystretch}{1.10}
\begin{tabular}[t]{lccc}
\hline \hline
                             & \textit{aG'g} & \textit{gG'a} & \textit{g'G'g} \\
\hline                                   
no. of transitions           &  2013         &  1401         &  1198          \\
6$-$26~GHz                   &    82$^{a,b}$ &    41$^a$     &    41$^a$      \\
38$-$70~GHz                  &   256         &   115         &   211          \\
200$-$230~GHz                &   553         &   280         &   177          \\
297$-$400~GHz                &  1122         &   965         &   769          \\
no. of $R$-branch trans.     &  1436         &  1338         &   950          \\
no. of $Q$-branch trans.     &   577         &    63         &   248          \\
no. of $a$-types             &   756         &  1372         &     6$^a$      \\
no. of $b$-types             &  1261         &    14$^a$     &   809          \\
no. of $c$-types             &    16$^{a,b}$ &    15$^a$     &   383          \\
no. of different lines       &  1176         &   723         &   718          \\
rms error of the fit         &  0.975        &  1.022        &  0.985         \\
standard deviation$^c$ (kHz) &   20.8        &   20.9        &   23.6         \\
uncertainty range (kHz)      &  10$-$90      &  10$-$50      &  10$-$70       \\
$J_{\rm max}$                &    69         &    71         &     69         \\
$K_{a{\rm , max}}$           &    38         &    45         &     33         \\

\hline
\end{tabular}\\[2pt]
\end{center}
$^a$ From Lovas et al. (2009).\\
$^b$ From Lockley et al. (2002).\\
$^c$ Given for completeness; see Sect.~\ref{results}.\\
\end{table}

Line overlap or proximity of two lines occurred quite rarely at these frequencies, therefore the uncertainties 
were estimated based on the base line quality and the signal-to-noise ratio. Assigned uncertainties 
at these low frequencies ranged mostly from 10 to 25~kHz, with some weak lines having uncertainties 
up to 50~kHz. The assigned uncertainties increased at higher frequencies up to 90~kHz. After each round 
of assignments, the need for more spectroscopic parameters was tested. We searched for the parameter 
that reduced the rms error of the fit most among the parameters that were useful based on the
previously employed parameters.   An additional parameter whose inclusion lead to a substantial reduction of the 
rms error was generally determined with great significance, meaning that its uncertainty was lower than 
one fifth of its magnitude. The search for additional parameters was continued as long as substantial 
reduction of the rms error was obtained.

\begin{figure}
\centering
  \includegraphics[width=9cm]{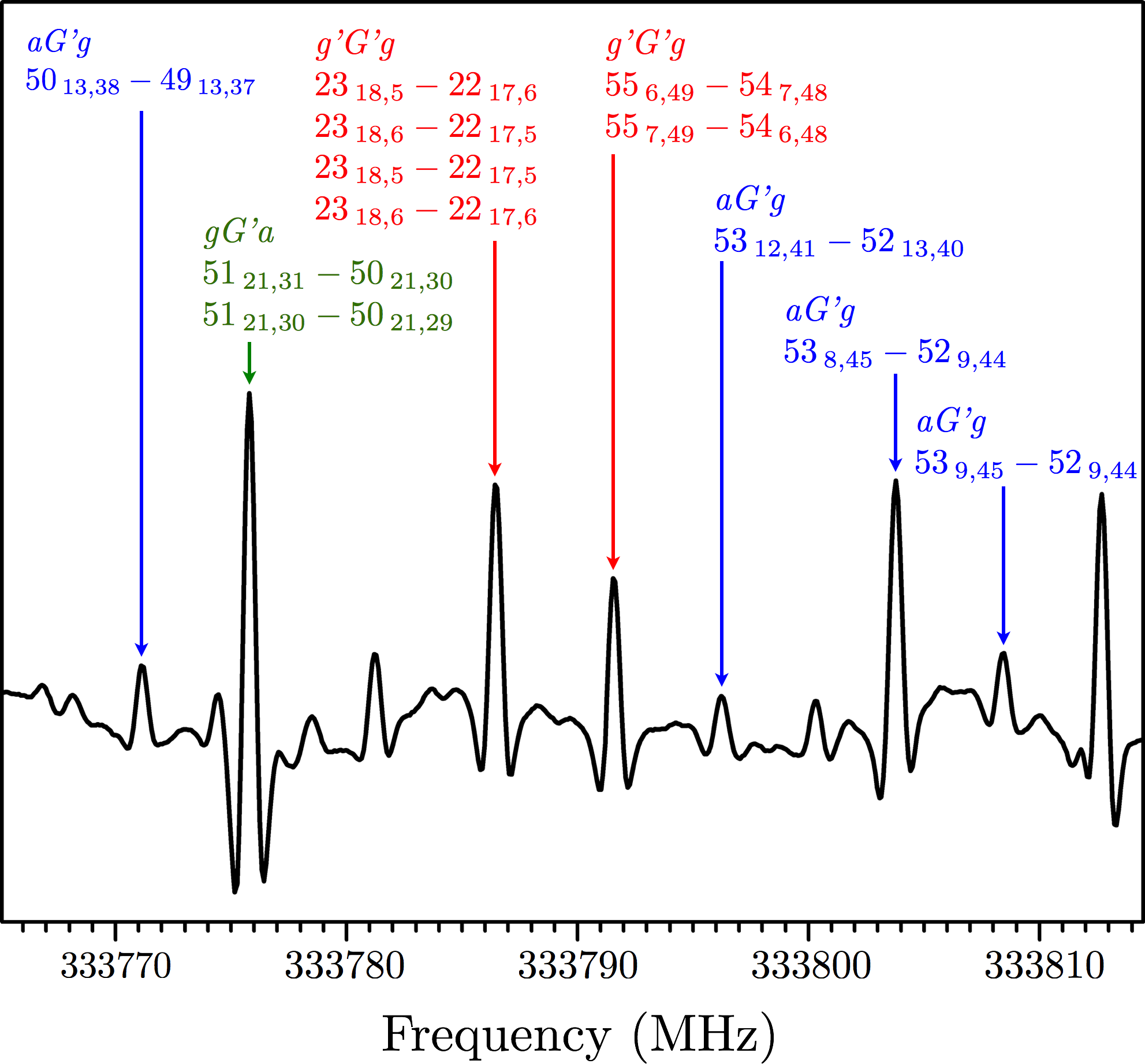} 
      \caption{Section of the rotational spectrum of 1,2-propanediol near 333.8~GHz. Transitions assignable 
               to  the three lowest energy conformers are marked. The $53_{8\,45} - 52_{8\,44}$ transition of 
               the $aG'g$ conformer occurs at 333812.0~MHz, but it is overlapped by a stronger, unidentified line. 
               All other unmarked lines belong to higher energy conformers or to excited states of the three 
               low-energy conformers. The transitions appear as approximate second derivatives of a 
               Gaussian line shape because of the 2$f$-modulation.}
\label{fig2}
\end{figure}

Subsequently, analyses of spectra in the 200$-$230~GHz region and then in the 297$-$400~GHz region were 
made in a similar way. Line overlap or proximity of lines was more widespread at higher frequencies 
and restricted the assignments somewhat in the 200$-$230~GHz and severely in the 297$-$400~GHz regions, in particular for weaker lines. Figure~\ref{fig2} shows a section of the spectrum near 334~GHz. 
Overlapping lines were not used in the fit except for unresolved asymmetry splitting. This refers to two 
$a$-, $b$-, or $c$-type transitions with the same $J$ quantum numbers in the upper and the lower state, 
the same $K_a$ (prolate pairing) or $K_c$ (oblate pairing), and $K_c$ or $K_a$ differing by one 
(note: $K_a + K_c = J$ or $J + 1$). The two transitions have the same intensity, and the frequency average 
is identical to the unsplit line center. For the \textit{aG'g} conformer with $a$- and $b$-dipole 
moment components of similar magnitude, oblate pairing may involve four transitions with similar intensities, 
two $a$- and two $b$-types. Analogously, prolate pairing of transitions of the \textit{g'G'g} conformer 
may involve four ($b$- and $c$-type) transitions with similar intensities. 
As the average frequency is in this special case identical to the intensity-weighted average, 
we omitted the intensity-weighting (specified in the line file after the uncertainties in SPFIT). 
The maximum $J$ quantum numbers of transitions used in the fits are 69, 71, and 69 for the \textit{aG'g}, 
\textit{gG'a}, and \textit{g'G'g}, respectively; the corresponding $K_a$ values are 38, 45, and 33.


Previous data from Lovas et al. (2009) were used in the final fits with the reported uncertainties 
with the exception of two lines of the \textit{aG'g} conformer and of three lines of the \textit{g'G'g} conformer, 
for which the uncertainties were increased from 2 to 5~kHz (four lines) and from 5 to 10~kHz (one line) because 
of high residuals in the final fits. In addition, three lines of the \textit{g'G'g} conformer were omitted 
for the same reason. Uncertainties of 4~kHz were assigned to the \textit{aG'g} data from 
Lockley et al. (2002). The \textit{gG'a} data from Caminati (1981) were not 
used in Lovas et al. (2009), and we did not use them either. 
Ultimately, a full set of up to sextic centrifugal distortion parameters plus one octic term were determined 
for the lowest energy conformer; two parameters less each were used in the final fits of the two higher 
energy conformers. The rms error of each fit, overall as well as restricted to one of the three frequency 
windows or to the previous data sets, was close to 1.0 throughout, in most cases slightly lower. 
Table~\ref{statistics} presents the unitless rms error as well as the standard deviation of each fit along 
with additional statistical information. The standard deviation is only given for completeness. It cannot 
be used as a measure for the quality of the fit because we used more than one uncertainty to 
reflect the changing quality of the experimental lines.  
The final parameters of the three lowest energy conformers of 1,2-propanediol are given in 
Table~\ref{spec-parameter}.


\begin{table*}
\begin{center}
\caption{Spectroscopic parameters$^{a}$ (MHz) of the three lowest energy conformers of 1,2-propanediol.}
\label{spec-parameter}
\renewcommand{\arraystretch}{1.10}
\begin{tabular}[t]{l@{}lr@{}lr@{}lr@{}l}
\hline \hline
\multicolumn{2}{l}{Parameter} & \multicolumn{2}{c}{\textit{aG'g}} & \multicolumn{2}{c}{\textit{gG'a}} & \multicolumn{2}{c}{\textit{g'G'g}} \\
\hline
$A$      &                   &   8572&.057374\,(95)    &   8393&.400915\,(198)   &   8536&.771860\,(108)   \\
$B$      &                   &   3640&.099657\,(37)    &   3648&.559609\,(45)    &   3604&.192077\,(50)    \\
$C$      &                   &   2790&.972773\,(33)    &   2778&.302133\,(44)    &   2778&.337515\,(52)    \\
$D_K$    & $\times 10^{3}$   &      1&.863902\,(170)   &      2&.38752\,(275)    &      1&.862365\,(315)   \\
$D_{JK}$ & $\times 10^{3}$   &      6&.085380\,(133)   &      5&.337883\,(204)   &      6&.107365\,(207)   \\
$D_{J}$  & $\times 10^{3}$   &      0&.6116173\,(186)  &      0&.6593872\,(172)  &      0&.6178250\,(226)  \\
$d_{1}$  & $\times 10^6$     & $-$163&.0629\,(72)      & $-$182&.78504\,(284)    & $-$157&.1335\,(44)      \\
$d_{2}$  & $\times 10^6$     &  $-$62&.7055\,(46)      &  $-$61&.6642\,(103)     &  $-$60&.0612\,(70)      \\
$H_{K}$  & $\times 10^9$     &     69&.320\,(103)      &       &                 &     70&.809\,(222)      \\
$H_{KJ}$ & $\times 10^9$     &  $-$43&.356\,(153)      &  $-$30&.218\,(183)      &  $-$43&.284\,(211)      \\
$H_{JK}$ & $\times 10^9$     &  $-$12&.9991\,(310)     &  $-$13&.913\,(37)       &  $-$12&.932\,(104)      \\
$H_J$    & $\times 10^{9}$   &      0&.1462\,(33)      &      0&.18016\,(230)    &      0&.1637\,(37)      \\
$h_{1}$  & $\times 10^{12}$  &  $-$10&.21\,(169)       &       &                 &       &                 \\
$h_{2}$  & $\times 10^{12}$  &  $-$73&.42\,(118)       &  $-$85&.75\,(171)       &  $-$78&.60\,(217)       \\
$h_{3}$  & $\times 10^{12}$  &     25&.05\,(36)        &     21&.79\,(69)        &     22&.29\,(75)        \\
$L_{JK}$ & $\times 10^{12}$  &   $-$0&.2276\,(274)     &     $-$0&.4312\,(309)   &       &                 \\

\hline
\end{tabular}\\[2pt]
\end{center}
$^a$ Watson's $S$-reduction was used in the representation $I^r$. Numbers in parentheses are one standard deviation 
in units of the least significant digits.\\
\end{table*}

The newly recorded transitions with their assignments, uncertainties, and residuals between 
observed frequency and those calculated from the final set of spectroscopic parameters are available 
in the supplementary material, as outlined in the appendix. The entire line, parameter, and fit files 
along with additional auxiliary files will be available in the spectroscopy 
section\footnote{Internet address: http://www.astro.uni-koeln.de/cdms/daten}  
of the CDMS~\citep{CDMS_1,CDMS_2}. Updated predictions for all three lower energy conformers of 
1,2-propanediol will be available in the catalog section of the CDMS.


\section{Discussion and conclusion}
\label{discussion}


\begin{figure}
\centering
  \includegraphics[width=9cm]{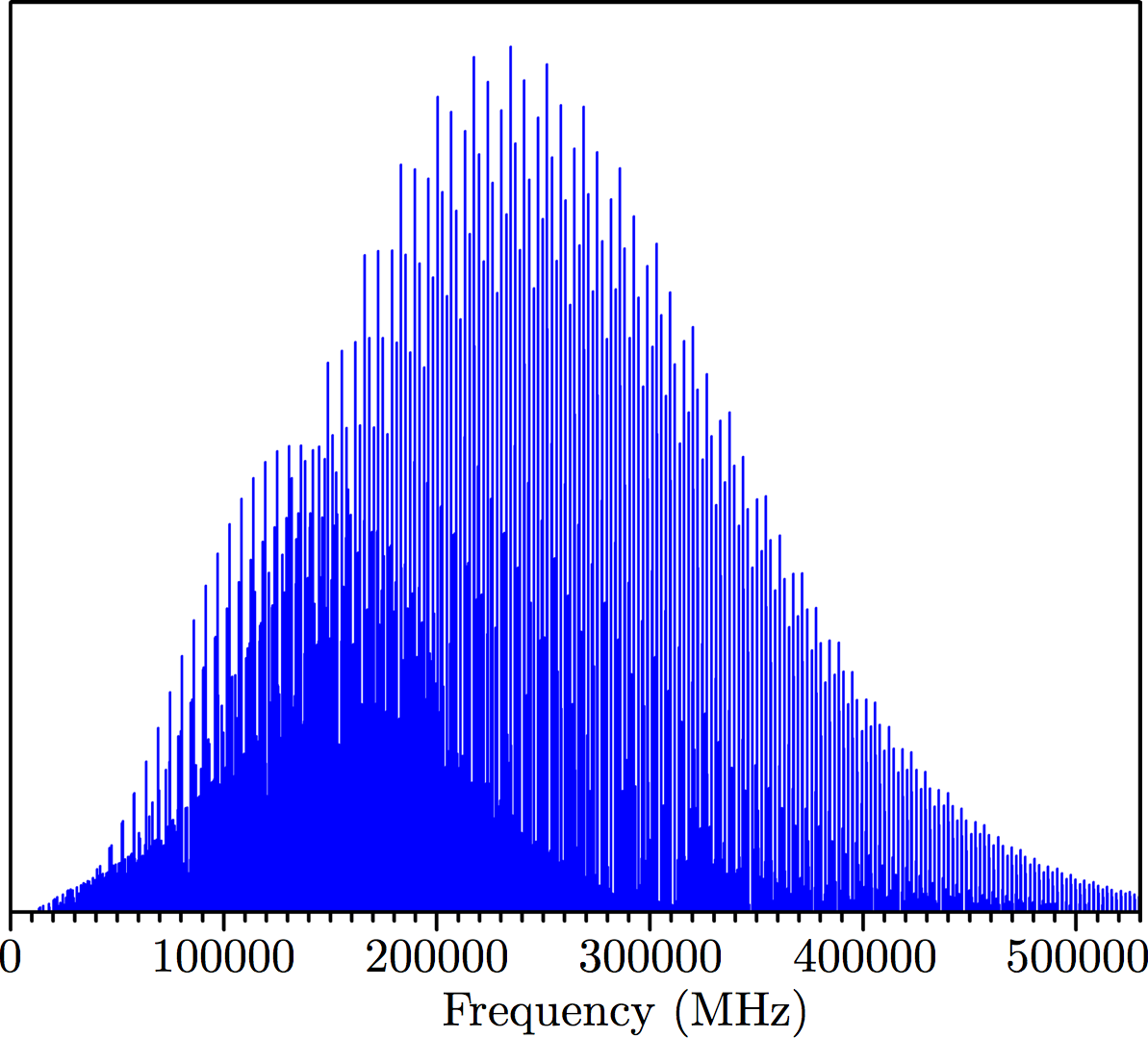} 
      \caption{Stick spectrum of the lowest energy \textit{aG'g} conformer of propanediol at 50~K.}
\label{stick-spectrum}
\end{figure}


The rotational constants of all three 1,2-propanediol conformers are very similar, as one can see in 
Table~\ref{spec-parameter}. Therefore, it is not surprising that the centrifugal distortion parameters 
are mostly very similar as well. A rotational temperature of 50~K was recently reported for the lighter 
homologue ethanediol in Sgr~B2(N) \citep{survey_SgrB2N-M_2013}. Assuming that the rotational temperature 
of propanediol in this source will be similar, the quantum number range covered for 1,2-propanediol 
in the present work is sufficient for astronomical observations, see Fig.~\ref{stick-spectrum}. Some 
extrapolation to higher quantum numbers is possible in case of much higher rotational temperatures. 
We expect the actual transition frequency to be close to the prediction, that is, within three to five 
times the predicted uncertainty, as long as the predicted uncertainty does not exceed 300~kHz.

According to Belloche et al. (2013), it is advantageous to search for the weak spectroscopic features 
of a rather complex molecule at frequencies below the Boltzmann peak at an expected rotational temperature. 
At centimeter wavelengths, however, local thermodynamic equilibrium (LTE) with one rotational temperature 
may be a poor assumption even for a prolific hot-core source such as Sgr~B2(N), at least in case 
of the lowest energy transitions. These transitions sample in particular the outer, less dense, and often 
colder envelope that surrounds the hot core. In fact, it was shown recently that the low-energy 
transitions of methyl formate, observed with the 100~m Greenbank telescope up to $\sim$50~GHz, deviated 
significantly from LTE, and several of these transitions were masing weakly \citep{MeFo_non-LTE_2014}.

An interesting aspect is the question whether we should expect an even greater molecular complexity to 
occur in the ISM. Bossa et al. (2009) reported on the thermal formation of aminomethanol 
in laboratory experiments consisting of H$_2$O, NH$_3$, and H$_2$CO. The molecule was identified 
in the solid phase by infrared spectroscopy. It is not known so far whether the molecule can be transferred 
into the gas phase without complete decomposition. Moreover, it is not known what its average lifetime 
in the gas phase would be compared with related molecules. Nevertheless, this intriguing result may 
indicate that methanediol, 1,1-ethanediol, 1,1-propanediol, 2,2-propanediol, etc. could also be formed 
in interstellar ice analogs. In this context, it may be of interest that carbonic acid, OC(OH)$_2$, 
is fairly stable in water-poor environments, in particular at low temperatures, and that it can be 
transferred into the gas phase without complete decomposition \cite{H2CO3_2013}. The microwave spectrum 
of carbonic acid was also recorded recently \cite{H2CO3_rot_2011}. And while OS(OH)$_2$ is, to our 
knowledge, not stable in the gas phase, O$_2$S(OH)$_2$, sulfuric acid, is, and its rotational spectrum 
was studied extensively quite recently \cite{H2SO4_rot_2013}. Apparently, the stability of a molecule 
containing two OH groups (or one OH group and one NH$_2$ group) attached to one atom X depends critically 
on the atom X and on other atoms or molecule groups attached to X.


\begin{acknowledgements}

These investigations have been supported by the Deutsche Forschungsgemeinschaft (DFG) in the framework 
of the collaborative research grant SFB~956, project B3. JBB is grateful for support from the Marie Curie
Intra-European Fellowship (FP7-PEOPLE-2011-IEF-299258).

\end{acknowledgements}


\section*{Appendix A. Supplementary material}
\label{Appendix}

The newly recorded experimental transition frequencies for the \textit{aG'g}, 
\textit{gG'a}, and \textit{g'G'g} conformers of 1,2-propanediol are given as 
Tables~\ref{supplement}, 5 and 6, respectively, in the supplementary 
material as text files. Only the first and the last ten lines of 
Table~\ref{supplement} appear in the paper edition. 
The tables give the rotational quantum numbers $J$, $K_a$, and $K_c$ 
for the upper state followed by those for the lower state. The observed 
transition frequency is given in megahertz units with its uncertainty 
and the residual between observed frequency and that calculated from 
the final set of spectroscopic parameters. In the case of unresolved 
asymmetry splitting, frequencies and residuals refer to the unsplit 
line center.

Additional text files {\tt aG'g.txt}, {\tt gG'a.txt}, and {\tt g'G'g.txt} 
provide the internal coordinates of the three 
propanediol conformers. A schematic representation is given in Table~\ref{z-matrix}.


\begin{table*}[ht]
\begin{center}
\caption{Transitions of the \textit{aG'g} conformer of 1,2-propanediol, observed transition frequencies (MHz), experimental uncertainties 
Unc. (MHz), and residuals o$-$c between observed frequency and that calculated from 
the final set of spectroscopic parameters.}
\label{supplement}
\begin{tabular}{rrrrrrr@{}lrr}
\hline \hline
$J'$ & $K_a'$ & $K_c'$  & $J''$ & $K_a''$ & $K_c''$ & \multicolumn{2}{c}{Frequency} & Unc. & o$-$c \\
\hline

  1&  0&  1&  0&  0&  0&                     6431.&0705&   0.004 &    0.00052 \\
  2&  1&  1&  2&  0&  2&                     6730.&6486&   0.004 & $-$0.00132 \\
  5&  1&  5&  4&  2&  2&                     7764.&4042&   0.004 &    0.00652 \\
  2&  0&  2&  1&  1&  1&                     7829.&7009&   0.004 &    0.00263 \\
  3&  1&  2&  3&  0&  3&                     8330.&2346&   0.004 & $-$0.00261 \\
  4&  1&  3&  4&  1&  4&                     8439.&0989&   0.004 &    0.00071 \\
  4&  1&  3&  4&  0&  4&                    10733.&5960&   0.004 & $-$0.00131 \\
  1&  1&  1&  0&  0&  0&                    11363.&0101&   0.004 & $-$0.00422 \\
  2&  1&  2&  1&  1&  1&                    12012.&9795&   0.004 &    0.00022 \\
  3&  0&  3&  2&  1&  1&                    12169.&5787&   0.004 &    0.00079 \\
   &   &   &   &   &   &                          &    &         &            \\
 61& 27& 34& 60& 27& 33&                   397547.&6360&   0.030 & $-$0.04320 \\
 61& 27& 35& 60& 27& 34&                   397547.&6360&   0.030 & $-$0.04320 \\
 67&  5& 62& 66&  6& 61&                   397802.&5190&   0.055 &    0.01818 \\
 67&  5& 62& 66&  5& 61&                   397802.&5190&   0.055 &    0.01818 \\
 67&  6& 62& 66&  5& 61&                   397802.&5190&   0.055 &    0.01818 \\
 67&  6& 62& 66&  6& 61&                   397802.&5190&   0.055 &    0.01818 \\
 68&  4& 64& 67&  5& 63&                   398920.&0610&   0.080 & $-$0.01838 \\
 68&  4& 64& 67&  4& 63&                   398920.&0610&   0.080 & $-$0.01838 \\
 68&  5& 64& 67&  4& 63&                   398920.&0610&   0.080 & $-$0.01838 \\
 68&  5& 64& 67&  5& 63&                   398920.&0610&   0.080 & $-$0.01838 \\

\hline
\end{tabular}\\[2pt]
\end{center}
  Notes. This table as well as those of other conformers are
  available in their entirety in the electronic edition in the online journal: 
  http://cdsarc.ustrasbg.fr/cgi-bin/VizieR?-source=J/A+A/Vol/Num.  
  A portion is shown here for guidance regarding its form and content.\\
\end{table*}

\setcounter{table}{6}

\begin{table*}[ht]
\begin{center}
\caption{Part of the schematic representation of the internal coordinates (in 100~pm and degrees) 
         of the \textit{aG'g} conformer of 1,2-propanediol given in {\tt aG'g.txt}.}
\label{z-matrix}
\begin{tabular}{ccccccc}
\hline \hline
atom & atom ref. no. & bond length label & atom ref. no. & bond angle label & atom ref. no. & dihedral angle label \\
\hline

 H    &    &     &   &     &   &    \\
 O    & 1  & B1  &   &     &   &    \\
 C    & 2  & B2  & 1 & A1  &   &    \\
 C    & 3  & B3  & 2 & A2  & 1 & D1 \\
      &    &     &   &     &   &    \\
label & \multicolumn{6}{r}{value}   \\
\hline
   B1 & \multicolumn{6}{r}{0.96170000} \\
   B2 & \multicolumn{6}{r}{1.43000000} \\
      &    &     &   &     &   &    \\
   D9 & \multicolumn{6}{r}{178.50000000} \\
   D10& \multicolumn{6}{r}{59.30000000} \\

\hline
\end{tabular}\\[2pt]
\end{center}
  Notes. This text file as well as those of other conformers are
  available in their entirety in the electronic edition in the online journal: 
  http://cdsarc.ustrasbg.fr/cgi-bin/VizieR?-source=J/A+A/Vol/Num.  
  A portion is shown here for guidance regarding its form and content.\\
\end{table*}

\end{document}